\documentclass[iop,superscriptaddress,preprintnumbers,a4,showpacs,floatfix,twocolumn]{revtex4-1}  
\usepackage{amsmath, amssymb, psfrag, tabls, dcolumn, bm, color}
\usepackage[sort&compress]{natbib}
\usepackage{graphicx}
\usepackage{bm}

\begin{document}

\title{Bending-Induced Delamination of van der Waals Solids}

\author{Pekka Koskinen}
\email[email:]{pekka.koskinen@iki.fi}
\address{NanoScience Center, Department of Physics, University of Jyv\"askyl\"a, 40014 Jyv\"askyl\"a, Finland}

\pacs{61.46.-w,62.25.-g,68.65.Pq,68.55.-a}
% 68.65.Pq 	Graphene films
% 62.25.-g 	Mechanical properties of nanoscale systems
%68.35.Gy 	Mechanical properties; surface strains
%68.55.-a 	Thin film structure and morphology
%68.60.-p 	Physical properties of thin films, nonelectronic
%68.60.Bs 	Mechanical and acoustical properties 

%68.65.-k 	Low-dimensional, mesoscopic, nanoscale and other related systems: structure and nonelectronic properties
%61.46.-w 	Structure of nanoscale materials

\begin{abstract}
Although sheets of layered van der Waals solids offer great opportunities to custom-design nanomaterial properties, their weak interlayer adhesion challenges structural stability against mechanical deformation. Here, bending-induced delamination of multilayer sheets is investigated by molecular dynamics simulations, using graphene as an archetypal van der Waals solid. Simulations show that delamination of a graphene sheet occurs when its radius of curvature decreases roughly below $R_c=5.3\text{ nm}\times (\text{number of layers})^{3/2}$ and that, as a rule, one-third of the layers get delaminated. These clear results are explained by a general and transparent model, a useful future reference for guiding the design of nanostructured van der Waals solids.

\end{abstract}
\maketitle

\section{Introduction}
The richness of layered van der Waals solids such as graphene, hexagonal boron nitride, layered transition metal dichalgonides, and others, arises largely from the possibility to custom-design their properties.\cite{butler_AN_13, bjorkman_PRL_12} The electronic and electromechanical properties of each layer are unique\cite{hod_NL_09,koskinen_APL_11,johari_AN_12} and they remain so upon layering because of the weakness of chemical bonding.\cite{olsen_PRL_11,bjorkman_PRL_12} Therefore, a control over the stacking order of different layers enables the fabrication of sheets with custom-designed properties useful in electronic components, nanoelectromechanical devices, and flexible electronics.\cite{wang_nnano_12,butler_AN_13} 

%Flexible components are enabled by materials' highly elastic behaviour

%Recently applications with flexible electronics and electromechanical components have received particular attention.

In layered materials, however, flexibility with pertinent mechanical deformations may not be possible due to structural instability. Can the weak binding hold the sheet together upon deformation, or will it get delaminated? When is stability limited by critical tensile strain instead of interlayer adhesion? Insights to the underlying physics would help to guide the manipulation and the design of mechanics and structure-function relationships for these layered materials.

Because sheets of layered materials are thin, the primary deformation mode is bending. Bending-induced delamination of thin sheets is familiar even from mundane objects, such as cardboard boxes. At the nanoscale, a recent work showed direct evidence of bending-induced delamination in graphene and boron nitride nanoribbons.\cite{nikiforov_PRL_12} The work was supplemented by a model assuming that the delamination is triggered by constant surface strain. Somewhat related phenomenon is the rippling of bent tubes, observed in bent multi-walled carbon nanotubes.\cite{poncharal_science_99,mahadevan_EPL_04,Nikiforov2010} Other types of nanosheet detachments have been investigated in various contexts, especially in the presence of flat or corrugated substrates.\cite{kusminskiy_PRB_11,bunch_SSC_12,wagner_APL_12,Cranford2013} However, a comprehensive picture of the delamination of sheets of van der Waals solids is still lacking.

In this article I present results of bending-induced delamination of multilayer sheets from classical molecular dynamics simulations. The simulations were designed to answer the simple but fundamental questions: How much does an $N$-layer sheet tolerate bending before it delaminates? How many layers get delaminated? What is the precise mechanism that triggers delamination, and can it be modeled? In this article these questions, as it will turn out, all receive explicit answers.

\begin{figure}[t]
\includegraphics[width=8cm]{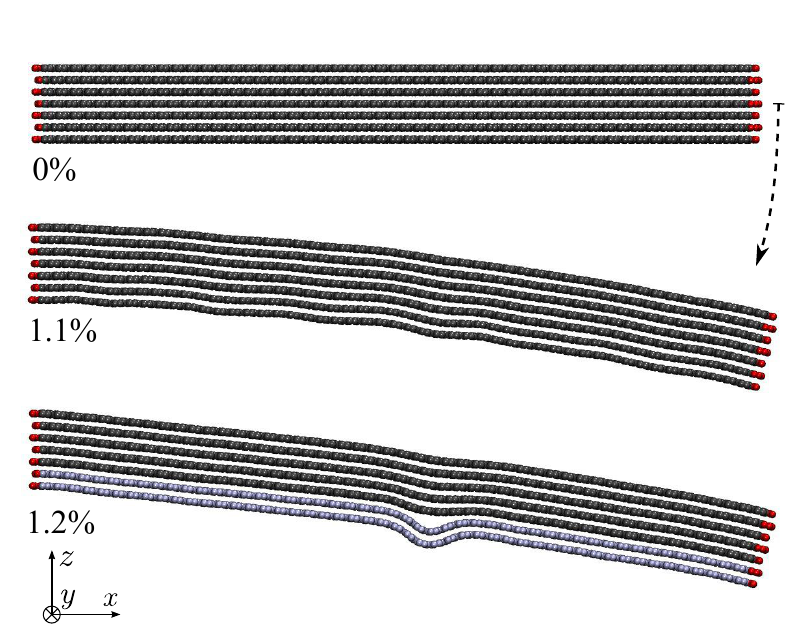}
\caption{(Color online) Delamination of multilayer graphene sheet upon bending. In seven-layer sheet two layers delaminate simultaneously when the radius of curvature decreases below $64$~nm, equivalent to $1.16$~\%\ surface strain (the numbers shown). Sheet is periodic in $y$-direction.}
\label{fig:setup}
\end{figure}

\section{Delamination in molecular dynamics simulations}
To simulate the bending-induced delamination, multilayer graphene sheets of finite length and infinite width were bent by introducing a constrained movement of the atoms in the ends (Fig.\ref{fig:setup}), while simulating the unconstrained atoms in the middle by a thermostat (see appendix). The degree of bending is quantified by the dimensionless parameter $\Theta=H/(2R)$, where $R$ is the radius of curvature of the neutral surface, $H=h(N-1)$ is the sheet thickness, $h$ is the interlayer distance, and $N$ is the number of layers. The parameter equals conveniently the tensile strain at the outermost layer and the compressive strain at the innermost layer, $\Theta=\varepsilon_{in}=-\varepsilon_{out}$.\cite{malola_PRB_08} The simulation constraints assumed harmonic bonds, which was a fair approximation because $\Theta\sim 1$~\%.\cite{koskinen_PRB_12} 

Simulations proceeded by increasing $\Theta$ from zero at a constant rate. Initially the sheets bent like a rigid plate, outer layers stretching, inner layers compressing. Upon further increasing $\Theta$, the inner layers developed delocalized undulations (view Fig.\ref{fig:setup} at an angle). At the critical value $\Theta=\Theta_c$ these undulations began to localize, inner layers slid with respect to the outer layers, resulting in delamination with an ever growing bump. After delamination the simulation was stopped because post-delamination events would have depended on the simulation constraints. The simulations were performed for thicknesses $N=2-14$ at $2$ K, $10$ K, $20$ K, $100$ K, and $300$ K temperatures, and repeated five times to obtain thermal variations.  

\begin{figure}[t!]
\includegraphics[width=9cm]{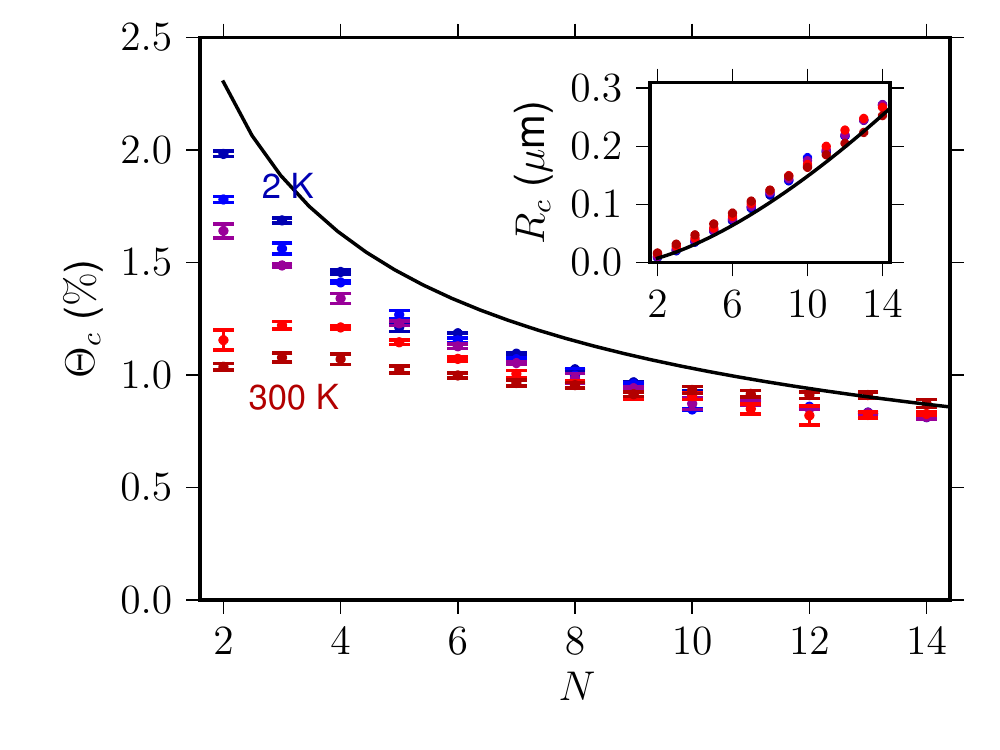}
\caption{(Color online) Critical surface strains for delamination as a function of sheet thickness. Color codes, temperatures of $2$ K (dark blue), $10$ K, $20$ K, $100$ K, and $300$ K (dark red); error bars, deviations from five simulations for each system; solid line, model prediction from Eq.(\ref{eq:thetac}). Inset: Corresponding critical radii of curvature as a function of sheet thickness.}
\label{fig:Theta}
\end{figure}

The first central result is the dependence of $\Theta_c$ on sheet thickness (Fig.\ref{fig:Theta}). For low temperatures $\Theta_c$ decreases monotonically with increasing $N$. At higher temperatures the dependence is non-monotonic because the shapes of the thinnest sheets were prone to fluctuate and create temporary excess curvature on top of the constrained one. When $N$ increases, however, this effect vanishes and the effect of temperature diminishes. The values of $\Theta_c=H/2R_c$ convert into critical radii of curvature $R_c$ in the sub-$\mu$m range, large enough to be taken seriously in device fabrication (inset of Fig.\ref{fig:Theta}). 

%The delamination occurs at the point where the strain energy  (tensile stress at the outer surface and the compressive stress at the inner surface) becomes so large that it becomes beneficiable to create a bump. 

%The bump costs bending energy and loss of vdW energy, but it can release some of the expensive strain energy. 

%The effect of temperature is non-trivial. For thin sheets the temperature induces earlier delamination, because bending makes the sheet to viggle and create larger local radii of curvature already earlier. 

The second central result is that, as a rule, one-third of the layers delaminate (Fig.\ref{fig:layers}). Apart from thermal fluctuation higher temperatures, the rule is unexpectedly robust. The one-third rule is particularly important because it is independent of material parameters, as will be discussed below.

Even though the simulations were stopped after the delamination, the question of what happens afterwards is important from a practical aspect. If bending was simply continued beyond $\Theta_c$, delamination continued to propagate towards the layers outside. However, these post-delamination events depend on the sheet length and on the choice of external constraints. Since these choices lack proper motivation, I discuss here the initial delamination event alone. 

\section{Delamination captured by an analytical model}
To analyze the delamination mechanism and the dependence on material parameters, let us next develop a continuum elasticity model. 

Consider an $N$-layer sheet bent to a radius of curvature $R$. Prior to delamination, the strain in each layer is $\varepsilon_i=\Theta[2i/(N-1)-1]$, where the indexing $i=0,\ldots,N-1$ starts from the innermost layer. When delamination begins, the innermost $N_d$ layers displace inwards, making a bump that I model by the profile $y(x)=d\cos^2(x\pi /w)$, where $d$ is bump height, $w$ is bump width, and $x$ ($\left|x\right|<w/2$) is the distance from bump center along the arc. In what follows, I analyze the delamination by inspecting how different energies contribute to the creation of such a bump. 

\begin{figure}[t!]
\includegraphics[width=9cm]{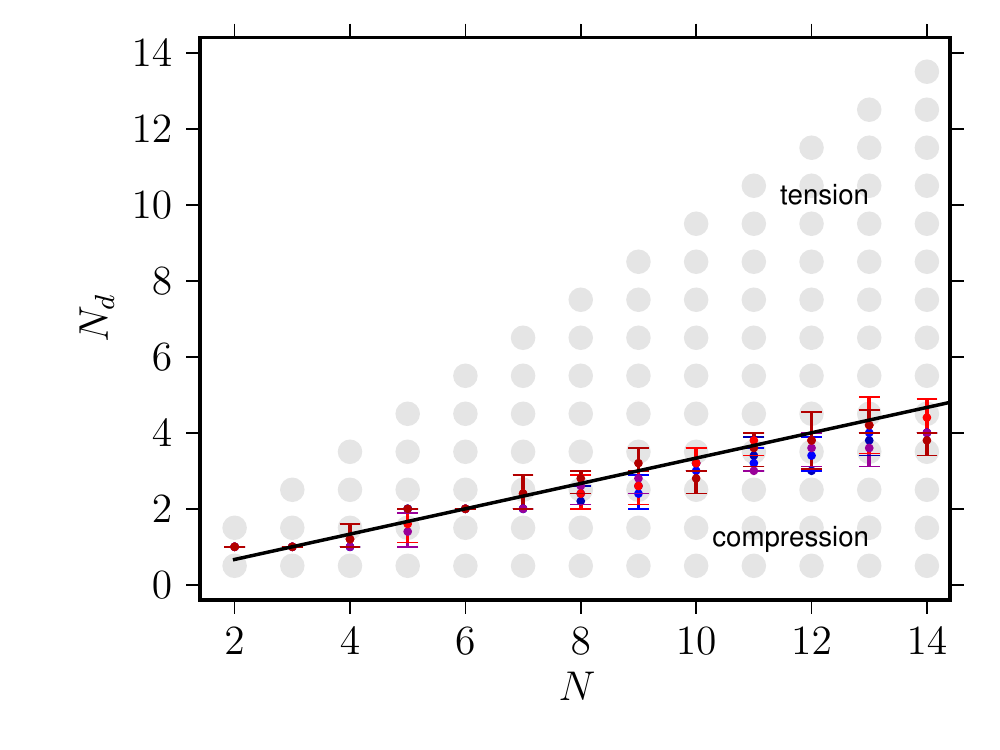}
\caption{(Color online) Number of delaminated layers as a function of sheet thickness. Color codes, temperatures of $2$ K (dark blue), $10$ K, $20$ K, $100$ K, and $300$ K (dark red); error bars, deviations from five simulations for each system; solid line, model prediction ($N_d=N/3$).}
\label{fig:layers}
\end{figure}

The first energy contribution is from strain. The stress in each layer is $k\varepsilon_i$, where $k$ is the in-plane modulus. The stress is constant during the initial phases of the delamination because the relatively long layers slide easily with respect to one another, and also because we investigate the limit of emerging bump, that is, small $d$.\cite{zheng_PRL_08,savini_carbon_11} Thus, a length change $\Delta l$ under constant stress makes up a change $k\varepsilon_i \Delta l$ in strain energy. Using the bump profile above, the length change for the $N_d$ innermost layers is $\Delta l_b = \pi^2d^2/4w$, which yields an energy change $k\Delta l_b\sum_{i=0}^{N_d-1} \varepsilon_i$. The contribution is negative because $\varepsilon_i<0$ inside. At the same time, bit more subtly, the lengths of the $N-N_d$ outermost layers comply to the sliding of the inner layers by shortening the corresponding amount and yielding an energy change $k(-\Delta l_b)\sum_{i=N_d}^{N-1} \varepsilon_i$. Adding the two terms, the strain energy upon bump formation changes by
\begin{equation}
E_s=-\frac{k \pi^2 \Theta N_d(N-N_d)}{2w(N-1)}d^2.
\label{eq:Es}
\end{equation}

The second energy contribution is from bending. The bending energy density for all $N_d$ layers is $\tfrac{1}{2}\kappa [y''(x)]^2$, where $\kappa$ is the bending modulus. Integrating over the bump profile, we get the total bending energy
\begin{equation}
E_B=\frac{N_d\kappa\pi^2}{w^3}d^2.
\label{eq:EB}
\end{equation}
Note that, apart from the expressions for $\varepsilon_i$, I ignore all other effects from the overall $R^{-2}$ curvature; this approximation improves upon increasing $R$ and $N$. 

The third energy contribution is from interlayer adhesion. Because it is reasonable to consider adhesion only between neighboring layers\cite{bjorkman_PRL_12}, energy changes only between the layers $i=N_d-1$ and $i=N_d$. Integrating the Lennard-Jones pair-potential $4 \epsilon [(h/r)^{12}-(h/r)^6]$ across two infinite sheets separated by $h'$ yields the interaction energy $V_{LJ}(h')=-2\pi \epsilon h^6 A_c^{-2} h'^{-4}[1-2/5(h/h')^6]$ with a minimum at $h'=h$, where $A_c$ is the surface area per atom. Assuming that this energy is local and integrating across the widening interlayer spacing of $h'=h+y$, we get to the lowest order in $d$
\begin{equation}
E_{adh}=\frac{9\pi\epsilon w}{A_c^2} d^2.
\label{eq:Eadh}
\end{equation} 

%\begin{align}
%\nonumber
%E=d^2[ -\frac{\Theta/2\pi^2 k N_d(N-N_d)}{w (N-1)}+&\\
%  \frac{N_d\kappa \pi^4}{w^3}+\frac{9\pi\epsilon w}{A_c^2} &].
%\end{align}

All three contributions (\ref{eq:Es}), (\ref{eq:EB}), and (\ref{eq:Eadh}) are quadratic in $d$, and the total cost to form the bump is
\begin{equation}
E=d^2\left[-\frac{\Theta k\pi^2 N_d(N-N_d)}{2w (N-1)}+
  \frac{N_d\kappa \pi^4}{w^3}+\frac{9\pi\epsilon w}{A_c^2}\right].
\label{eq:E}
\end{equation}
This expression offers a transparent stability analysis. When $\Theta$ is small and the coefficient of $d^2$ positive, the sheet is stable against delamination; the gain in strain energy is too small compared to the cost in bending and adhesion. When $\Theta$ increases the capacity to release strain energy grows until at $\Theta_c$ the coefficient of $d^2$ equals zero, making it possible to create the bump without cost, triggering the delamination. At $\Theta=\Theta_c$, the number of delaminated layers can be calculated by energy minimization from $\partial E/\partial N_d=0$, and the width of the bump from $\partial E/\partial w=0$. This gives us three equations out of which the unknowns ($\Theta_c$, $N_d$, and $w$) can be solved. 

The solution for $\Theta_c$ at the limit of large $N$ is simply 
\begin{equation}
\Theta_c\approx \frac{35\sqrt{\epsilon \kappa}}{A_c k}\frac{1}{\sqrt{N}},
\label{eq:thetac}
\end{equation}
which becomes $\Theta_c=0.032\times N^{-1/2}$ with graphene parameters, agreeing well with the low-temperature molecular dynamics simulations (Fig.\ref{fig:Theta}).\cite{footnote} Although residing close to $1.5$~\%\ for few-layer graphene, $\Theta_c$ is not quite constant, as assumed earlier.\cite{nikiforov_PRL_12} This expression infers a critical radius for stability $R_c\approx Nh/2\Theta_c=5.3\times N^{3/2}$~nm (inset of Fig.\ref{fig:Theta}). Delamination becomes easier when in-plane modulus increases or when adhesion and bending moduli decrease, which are plausible dependencies on material parameters.

The solution for $N_d$ is $N_d=N/3$, exactly. This one-third rule is particularly elegant because it is completely \emph{independent of material parameters}. The rule agrees with the simulations, where the number of delaminated layers is always an integer close to $N/3$, even though especially at higher temperatures sheets occasionally delaminated to different number of layers (Fig.\ref{fig:layers}). To a certain degree, the one-third rule can be understood by qualitative argumentation. Since the cost in adhesion is independent of $N_d$, it is governed by the competition between stretching and bending alone. Having $N_d=N/2$ would give the largest release in strain energy, but the strain in the layers near the neutral surface is so small that it becomes beneficial to retain their strain and to avoid the comparably large cost in bending energy. Hence, the best solution is $N_d$ between $N/2$ and $1$, and the value $N/3$ emerges from the functional form of bending and strain energies, independent of material parameters.

The model is expected to be universally valid, since the adhesion parameter in all van der Waals solids has the same nature and physical origin.\cite{bjorkman_PRL_12} The model is also unexpectedly successful especially in view of its simplicity. First, the model has no fitting parameters. Second, it takes into account the $1/R$ curvature only in the layer strains. Third, it ignores the undulations that were seen to precede the delamination. As expected, the agreement improves upon larger $N$ because some of the approximations get milder. In practice, the limit of (very) large $N$ is non-trivial, because the delamination itself possibly becomes less an instantaneous event, and more a dynamic process. Furthermore, the model ignores the clearly important entropic effects at higher temperatures. I remark that the solution for $w$, with the explicit expression
\begin{equation}
w = 1.13 (\kappa A_c^2 N/\varepsilon)^{1/4},
\end{equation}
is a by-product not to be compared with the simulations; it is defined only at the onset of delamination, and its comparison with the size of the post-delaminated bump would be vague. 

%The model could be improved to agree with the simulations better, but that would only complicate the analysis and compromise the model transparency; also the benefits of improvement would be limited. 

It is straightforward to generalize the model to heterogeneous multilayers. By considering layer-dependent bending and in-plane moduli, the strain energy becomes
\begin{equation}
E_s = \Delta l_b \left( \sum_{i=0}^{N_d-1} k_i \varepsilon_i - \sum_{i=N_d}^{N-1} k_i \varepsilon_i \right ),
\end{equation}
and the bending energy
\begin{equation}
E_B = \frac{\pi^2 d^2}{w^3} \sum_{i=0}^{N_d} \kappa_i.
\end{equation}
The adhesion energy remains equal to (\ref{eq:Eadh}) with the parameter $\epsilon$ replaced by the adhesion between layers $i=N_d-1$ and $i=N_d$, even though it seems that the adhesion for layered materials has almost a universal value of $\sim 20$~meV/\AA$^2$.\cite{bjorkman_PRL_12} Heterogeneous multilayers could introduce several solutions because the varying material parameters could make the total energy a non-monotonous function of $N_d$. 

\section{Discussion and conclusion}
The implications of delamination are multifaceted. It has profound effect on the nanoelectromechanical properties of multilayer sheets, either through the delamination itself or through a more radical structural transformation triggered by the delamination. The delaminated structure itself would abruptly change the bending modulus of the sheet. Further, with its disturbed interlayer coupling, the bump would cause electronic scattering and otherwise modify sheet's electronic properties. The properties of multilayer graphene, namely, depend very strongly on the number of layers.\cite{DasSarma2011} Regarding structural stability in general, Eq.(\ref{eq:thetac}) gives a useful criterion: when $\Theta_c$ approaches the maximum tensile strain of the layered material, bending would cause fracture instead of delamination. Pristine graphene withstands tensile strain more than $\sim 15$~\%\ ($\gg 2$~\%), so delamination occurs before fracture.\cite{wei_nmat_12} 

%Depending on the experimental constraints, 
Delamination could also be a trigger for more radical structural transformation. It could launch the formation of standing collapsed wrinkles, protrusions emerging from the sheet created due to easy sliding.\cite{zhu_NL_12} It could even trigger exfoliation, possibly involving the intercalation of particles inside the cavity below the bump, a particularly plausible mechanism for narrow sheets and ribbons. This way the delamination would become irreversible. In fact, it is conceivable that such mechanism would be partly responsible for the exfoliation of dissolved graphene flakes under intense sonication. Sonication would excite bending of the flakes, thereby causing delamination and easier intercalation of the particles from solution between the layered structure.\cite{duong_nature_12,coleman_science_11} 

%#although influencing the properties of multilayer sheets in many ways, 

To conclude, simulations and modeling suggest that delamination is a robust and predictable phenomenon. Especially the one-third rule for the number of delaminated layers could be a valuable tool for the manipulation and design of multilayer sheets. Thus, the comprehensive understanding of the mechanism and the powerfully simple results for the critical curvatures is likely to become a useful future reference for designing nanoelectromechanical devices of van der Waals solids.

\section*{Acknowledgements}
I acknowledge Tuomas Tallinen for discussions, the Academy of Finland for funding, and the Finnish IT Center for Science (CSC) for computational resources.

\appendix

\section{Methods}
%\emph{Methods ---} 
Simulations used the LAMMPS package with AIREBO potential for atomic interactions\cite{plimpton_JCP_95,stuart_JCP_00}; potential includes the Lennard-Jones $6-12$ potential for the interlayer adhesion (cutoff$=3\sigma$). The potential yields bending modulus of $\kappa=0.95$~eV, in-plane modulus of $k=21$~eV/\AA$^2$, and interlayer separation of $h=3.38$~\AA\ with $17$~meV/\AA$^2$ adhesion energy ($\epsilon=2.7$~meV). To model the structures, $N$ graphene layers of length $L_x=100\times N$~\AA\ were stacked in $xy$-plane with Bernal stacking (Fig.\ref{fig:setup}). The edges had zigzag profile and the length in $y$ direction was $L_y=2.5$~\AA\ with periodic boundary conditions (akin to a zigzag graphene nanoribbon in $y$-direction of width $L_x$; such a short periodic length was sufficient to account for the quasi-static motion in $y$-direction, as confirmed by simulations using much larger periodic lengths). After equilibrating $L_x$ at the respective remperature, atoms within $2.5$~\AA\ from the left edge were fixed, and bending was introduced by moving and turning the atoms within $2.5$~\AA\ from the right edge so as to create a perfect circular arc for the unconstrained atoms (treated with a Langevin thermostat at given temperature with $1$~ps relaxation time). Bending was increased at a constant rate $d\Theta/dt= 2.5\times 10^{-8}/N \text{ fs}^{-1}$, propagated using a $2$~fs time step. Bending was slow enough to be quasi-static, and delaminations were reversible. The simulation was stopped upon delamination because the subsequent events would have depended on $L_x$ and the imposed constraints. There would have been more elegant options to apply pure bending\cite{dumitrica_JMPS_07, koskinen_PRL_10}, but here the bending is pure enough for all practical purposes.

%\bibliography{library,notes}

\begin{thebibliography}{29}%
\makeatletter
\providecommand \@ifxundefined [1]{%
 \@ifx{#1\undefined}
}%
\providecommand \@ifnum [1]{%
 \ifnum #1\expandafter \@firstoftwo
 \else \expandafter \@secondoftwo
 \fi
}%
\providecommand \@ifx [1]{%
 \ifx #1\expandafter \@firstoftwo
 \else \expandafter \@secondoftwo
 \fi
}%
\providecommand \natexlab [1]{#1}%
\providecommand \enquote  [1]{``#1''}%
\providecommand \bibnamefont  [1]{#1}%
\providecommand \bibfnamefont [1]{#1}%
\providecommand \citenamefont [1]{#1}%
\providecommand \href@noop [0]{\@secondoftwo}%
\providecommand \href [0]{\begingroup \@sanitize@url \@href}%
\providecommand \@href[1]{\@@startlink{#1}\@@href}%
\providecommand \@@href[1]{\endgroup#1\@@endlink}%
\providecommand \@sanitize@url [0]{\catcode `\\12\catcode `\$12\catcode
  `\&12\catcode `\#12\catcode `\^12\catcode `\_12\catcode `\%12\relax}%
\providecommand \@@startlink[1]{}%
\providecommand \@@endlink[0]{}%
\providecommand \url  [0]{\begingroup\@sanitize@url \@url }%
\providecommand \@url [1]{\endgroup\@href {#1}{\urlprefix }}%
\providecommand \urlprefix  [0]{URL }%
\providecommand \Eprint [0]{\href }%
\@ifxundefined \urlstyle {%
  \providecommand \doi  [0]{\begingroup \@sanitize@url \@doi}%
  \providecommand \@doi [1]{\endgroup \@@startlink {\doibase
  #1}doi:\discretionary {}{}{}#1\@@endlink }%
}{%
  \providecommand \doi  [0]{doi:\discretionary{}{}{}\begingroup
  \urlstyle{rm}\Url }%
}%
\providecommand \doibase [0]{http://dx.doi.org/}%
\providecommand \Doi [0]{\begingroup \@sanitize@url \@Doi }%
\providecommand \@Doi  [1]{\endgroup\@@startlink{\doibase#1}\@@Doi}%
\providecommand \@@Doi [1]{#1\@@endlink}%
\providecommand \selectlanguage [0]{\@gobble}%
\providecommand \bibinfo  [0]{\@secondoftwo}%
\providecommand \bibfield  [0]{\@secondoftwo}%
\providecommand \translation [1]{[#1]}%
\providecommand \BibitemOpen [0]{}%
\providecommand \bibitemStop [0]{}%
\providecommand \bibitemNoStop [0]{.\EOS\space}%
\providecommand \EOS [0]{\spacefactor3000\relax}%
\providecommand \BibitemShut  [1]{\csname bibitem#1\endcsname}%
%</preamble>
\bibitem [{\citenamefont {Butler}\ \emph {et~al.}(2013)\citenamefont {Butler},
  \citenamefont {Hollen}, \citenamefont {Cao}, \citenamefont {Cui},
  \citenamefont {Gupta}, \citenamefont {Gutie}, \citenamefont {Heinz},
  \citenamefont {Hong}, \citenamefont {Huang}, \citenamefont {Ismach},
  \citenamefont {Johnston-halperin}, \citenamefont {Kuno}, \citenamefont
  {Plashnitsa}, \citenamefont {Robinson}, \citenamefont {Ruoff}, \citenamefont
  {Salahuddin}, \citenamefont {Shan}, \citenamefont {Shi}, \citenamefont
  {Spencer}, \citenamefont {Terrones}, \citenamefont {Windl},\ and\
  \citenamefont {Goldberger}}]{butler_AN_13}%
  \BibitemOpen
  \bibfield  {author} {\bibinfo {author} {\bibfnamefont {S.~Z.}\ \bibnamefont
  {Butler}}, \bibinfo {author} {\bibfnamefont {S.~M.}\ \bibnamefont {Hollen}},
  \bibinfo {author} {\bibfnamefont {L.}~\bibnamefont {Cao}}, \bibinfo {author}
  {\bibfnamefont {Y.}~\bibnamefont {Cui}}, \bibinfo {author} {\bibfnamefont
  {J.~A.}\ \bibnamefont {Gupta}}, \bibinfo {author} {\bibfnamefont {H.~R.}\
  \bibnamefont {Gutie}}, \bibinfo {author} {\bibfnamefont {T.~F.}\ \bibnamefont
  {Heinz}}, \bibinfo {author} {\bibfnamefont {S.~S.}\ \bibnamefont {Hong}},
  \bibinfo {author} {\bibfnamefont {J.}~\bibnamefont {Huang}}, \bibinfo
  {author} {\bibfnamefont {A.~F.}\ \bibnamefont {Ismach}}, \bibinfo {author}
  {\bibfnamefont {E.}~\bibnamefont {Johnston-halperin}}, \bibinfo {author}
  {\bibfnamefont {M.}~\bibnamefont {Kuno}}, \bibinfo {author} {\bibfnamefont
  {V.~V.}\ \bibnamefont {Plashnitsa}}, \bibinfo {author} {\bibfnamefont
  {R.~D.}\ \bibnamefont {Robinson}}, \bibinfo {author} {\bibfnamefont {R.~S.}\
  \bibnamefont {Ruoff}}, \bibinfo {author} {\bibfnamefont {S.}~\bibnamefont
  {Salahuddin}}, \bibinfo {author} {\bibfnamefont {J.}~\bibnamefont {Shan}},
  \bibinfo {author} {\bibfnamefont {L.}~\bibnamefont {Shi}}, \bibinfo {author}
  {\bibfnamefont {O.~M.~G.}\ \bibnamefont {Spencer}}, \bibinfo {author}
  {\bibfnamefont {M.}~\bibnamefont {Terrones}}, \bibinfo {author}
  {\bibfnamefont {W.}~\bibnamefont {Windl}}, \ and\ \bibinfo {author}
  {\bibfnamefont {J.~E.}\ \bibnamefont {Goldberger}},\ }\href@noop {}
  {\bibfield  {journal} {\bibinfo  {journal} {ACS Nano}} (\bibinfo {year}
  {2013})}\BibitemShut {NoStop}%
\bibitem [{\citenamefont {Bj\"{o}rkman}\ \emph {et~al.}(2012)\citenamefont
  {Bj\"{o}rkman}, \citenamefont {Gulans}, \citenamefont {Krasheninnikov},\ and\
  \citenamefont {Nieminen}}]{bjorkman_PRL_12}%
  \BibitemOpen
  \bibfield  {author} {\bibinfo {author} {\bibfnamefont {T.}~\bibnamefont
  {Bj\"{o}rkman}}, \bibinfo {author} {\bibfnamefont {A.}~\bibnamefont
  {Gulans}}, \bibinfo {author} {\bibfnamefont {A.}~\bibnamefont
  {Krasheninnikov}}, \ and\ \bibinfo {author} {\bibfnamefont {R.}~\bibnamefont
  {Nieminen}},\ }\Doi {10.1103/PhysRevLett.108.235502} {\bibfield  {journal}
  {\bibinfo  {journal} {Physical Review Letters},\ }\textbf {\bibinfo {volume}
  {108}},\ \bibinfo {pages} {235502} (\bibinfo {year} {2012})}\BibitemShut {NoStop}%
\bibitem [{\citenamefont {Hod}\ and\ \citenamefont
  {Scuseria}(2009)}]{hod_NL_09}%
  \BibitemOpen
  \bibfield  {author} {\bibinfo {author} {\bibfnamefont {O.}~\bibnamefont
  {Hod}}\ and\ \bibinfo {author} {\bibfnamefont {G.~E.}\ \bibnamefont
  {Scuseria}},\ }\Doi {10.1021/nl900913c} {\bibfield  {journal} {\bibinfo
  {journal} {Nano letters},\ }\textbf {\bibinfo {volume} {9}},\ \bibinfo
  {pages} {2619} (\bibinfo {year} {2009})}\BibitemShut {NoStop}%
\bibitem [{\citenamefont {Koskinen}(2011)}]{koskinen_APL_11}%
  \BibitemOpen
  \bibfield  {author} {\bibinfo {author} {\bibfnamefont {P.}~\bibnamefont
  {Koskinen}},\ }\href@noop {} {\bibfield  {journal} {\bibinfo  {journal}
  {Appl. Phys. Lett.},\ }\textbf {\bibinfo {volume} {99}},\ \bibinfo {pages}
  {13105} (\bibinfo {year} {2011})}\BibitemShut {NoStop}%
\bibitem [{\citenamefont {Johari}\ and\ \citenamefont
  {Shenoy}(2012)}]{johari_AN_12}%
  \BibitemOpen
  \bibfield  {author} {\bibinfo {author} {\bibfnamefont {P.}~\bibnamefont
  {Johari}}\ and\ \bibinfo {author} {\bibfnamefont {V.~B.}\ \bibnamefont
  {Shenoy}},\ }\href@noop {} {\bibfield  {journal} {\bibinfo  {journal} {ACS
  Nano},\ }\textbf {\bibinfo {volume} {6}},\ \bibinfo {pages} {5449} (\bibinfo
  {year} {2012})}\BibitemShut {NoStop}%
\bibitem [{\citenamefont {Olsen}\ \emph {et~al.}(2011)\citenamefont {Olsen},
  \citenamefont {Yan}, \citenamefont {Mortensen},\ and\ \citenamefont
  {Thygesen}}]{olsen_PRL_11}%
  \BibitemOpen
  \bibfield  {author} {\bibinfo {author} {\bibfnamefont {T.}~\bibnamefont
  {Olsen}}, \bibinfo {author} {\bibfnamefont {J.}~\bibnamefont {Yan}}, \bibinfo
  {author} {\bibfnamefont {J.~J.}\ \bibnamefont {Mortensen}}, \ and\ \bibinfo
  {author} {\bibfnamefont {K.~S.}\ \bibnamefont {Thygesen}},\ }\href@noop {}
  {\bibfield  {journal} {\bibinfo  {journal} {Phys. Rev. Lett.},\ }\textbf
  {\bibinfo {volume} {107}},\ \bibinfo {pages} {156401} (\bibinfo {year}
  {2011})}\BibitemShut {NoStop}%
\bibitem [{\citenamefont {Wang}\ \emph {et~al.}(2012)\citenamefont {Wang},
  \citenamefont {Kalantar-Zadeh}, \citenamefont {Kis}, \citenamefont
  {Coleman},\ and\ \citenamefont {Strano}}]{wang_nnano_12}%
  \BibitemOpen
  \bibfield  {author} {\bibinfo {author} {\bibfnamefont {Q.~H.}\ \bibnamefont
  {Wang}}, \bibinfo {author} {\bibfnamefont {K.}~\bibnamefont
  {Kalantar-Zadeh}}, \bibinfo {author} {\bibfnamefont {A.}~\bibnamefont {Kis}},
  \bibinfo {author} {\bibfnamefont {J.~N.}\ \bibnamefont {Coleman}}, \ and\
  \bibinfo {author} {\bibfnamefont {M.~S.}\ \bibnamefont {Strano}},\ }\Doi
  {10.1038/nnano.2012.193} {\bibfield  {journal} {\bibinfo  {journal} {Nature
  Nanotechnology},\ }\textbf {\bibinfo {volume} {7}},\ \bibinfo {pages} {699}
  (\bibinfo {year} {2012})}\BibitemShut
  {NoStop}%
\bibitem [{\citenamefont {Nikiforov}\ \emph {et~al.}(2012)\citenamefont
  {Nikiforov}, \citenamefont {Tang}, \citenamefont {Wei}, \citenamefont
  {Dumitricǎ},\ and\ \citenamefont {Golberg}}]{nikiforov_PRL_12}%
  \BibitemOpen
  \bibfield  {author} {\bibinfo {author} {\bibfnamefont {I.}~\bibnamefont
  {Nikiforov}}, \bibinfo {author} {\bibfnamefont {D.-M.}\ \bibnamefont {Tang}},
  \bibinfo {author} {\bibfnamefont {X.}~\bibnamefont {Wei}}, \bibinfo {author}
  {\bibfnamefont {T.}~\bibnamefont {Dumitricǎ}}, \ and\ \bibinfo {author}
  {\bibfnamefont {D.}~\bibnamefont {Golberg}},\ }\Doi
  {10.1103/PhysRevLett.109.025504} {\bibfield  {journal} {\bibinfo  {journal}
  {Physical Review Letters},\ }\textbf {\bibinfo {volume} {109}},\ \bibinfo
  {pages} {025504} (\bibinfo {year} {2012})}\BibitemShut {NoStop}%
\bibitem [{\citenamefont {Poncharal}\ \emph {et~al.}(1999)\citenamefont
  {Poncharal}, \citenamefont {Wand}, \citenamefont {Ugarte},\ and\
  \citenamefont {de~Heer}}]{poncharal_science_99}%
  \BibitemOpen
  \bibfield  {author} {\bibinfo {author} {\bibfnamefont {P.}~\bibnamefont
  {Poncharal}}, \bibinfo {author} {\bibfnamefont {Z.~L.}\ \bibnamefont {Wand}},
  \bibinfo {author} {\bibfnamefont {D.}~\bibnamefont {Ugarte}}, \ and\ \bibinfo
  {author} {\bibfnamefont {W.~A.}\ \bibnamefont {de~Heer}},\ }\href@noop {}
  {\bibfield  {journal} {\bibinfo  {journal} {Science},\ }\textbf {\bibinfo
  {volume} {283}},\ \bibinfo {pages} {1513} (\bibinfo {year}
  {1999})}\BibitemShut {NoStop}%
\bibitem [{\citenamefont {Mahadevan}\ \emph {et~al.}(2004)\citenamefont
  {Mahadevan}, \citenamefont {Bico},\ and\ \citenamefont
  {McKinley}}]{mahadevan_EPL_04}%
  \BibitemOpen
  \bibfield  {author} {\bibinfo {author} {\bibfnamefont {L.}~\bibnamefont
  {Mahadevan}}, \bibinfo {author} {\bibfnamefont {J.}~\bibnamefont {Bico}}, \
  and\ \bibinfo {author} {\bibfnamefont {G.}~\bibnamefont {McKinley}},\ }\Doi
  {10.1209/epl/i2003-10099-9} {\bibfield  {journal} {\bibinfo  {journal}
  {Europhysics Letters},\ }\textbf {\bibinfo {volume} {65}},\ \bibinfo {pages}
  {323} (\bibinfo {year} {2004})}\BibitemShut {NoStop}%
\bibitem [{\citenamefont {Nikiforov}\ \emph {et~al.}(2010)\citenamefont
  {Nikiforov}, \citenamefont {Zhang}, \citenamefont {James},\ and\
  \citenamefont {Dumitrică}}]{Nikiforov2010}%
  \BibitemOpen
  \bibfield  {author} {\bibinfo {author} {\bibfnamefont {I.}~\bibnamefont
  {Nikiforov}}, \bibinfo {author} {\bibfnamefont {D.-B.}\ \bibnamefont
  {Zhang}}, \bibinfo {author} {\bibfnamefont {R.~D.}\ \bibnamefont {James}}, \
  and\ \bibinfo {author} {\bibfnamefont {T.}~\bibnamefont {Dumitrică}},\
  }\Doi {10.1063/1.3368703} {\bibfield  {journal} {\bibinfo  {journal} {Applied
  Physics Letters},\ }\textbf {\bibinfo {volume} {96}},\ \bibinfo {pages}
  {123107} (\bibinfo {year} {2010})}\BibitemShut {NoStop}%
\bibitem [{\citenamefont {{Viola Kusminskiy}}\ \emph
  {et~al.}(2011)\citenamefont {{Viola Kusminskiy}}, \citenamefont {Campbell},
  \citenamefont {{Castro Neto}},\ and\ \citenamefont
  {Guinea}}]{kusminskiy_PRB_11}%
  \BibitemOpen
  \bibfield  {author} {\bibinfo {author} {\bibfnamefont {S.}~\bibnamefont
  {{Viola Kusminskiy}}}, \bibinfo {author} {\bibfnamefont {D.}~\bibnamefont
  {Campbell}}, \bibinfo {author} {\bibfnamefont {A.}~\bibnamefont {{Castro
  Neto}}}, \ and\ \bibinfo {author} {\bibfnamefont {F.}~\bibnamefont
  {Guinea}},\ }\Doi {10.1103/PhysRevB.83.165405} {\bibfield  {journal}
  {\bibinfo  {journal} {Physical Review B},\ }\textbf {\bibinfo {volume}
  {83}},\ \bibinfo {pages} {165405} (\bibinfo {year} {2011})}\BibitemShut {NoStop}%
\bibitem [{\citenamefont {Bunch}\ and\ \citenamefont
  {Dunn}(2012)}]{bunch_SSC_12}%
  \BibitemOpen
  \bibfield  {author} {\bibinfo {author} {\bibfnamefont {J.}~\bibnamefont
  {Bunch}}\ and\ \bibinfo {author} {\bibfnamefont {M.}~\bibnamefont {Dunn}},\
  }\Doi {10.1016/j.ssc.2012.04.029} {\bibfield  {journal} {\bibinfo  {journal}
  {Solid State Communications},\ }\textbf {\bibinfo {volume} {152}},\ \bibinfo
  {pages} {1359} (\bibinfo {year} {2012})}\BibitemShut {NoStop}%
\bibitem [{\citenamefont {Wagner}\ and\ \citenamefont
  {Vella}(2012)}]{wagner_APL_12}%
  \BibitemOpen
  \bibfield  {author} {\bibinfo {author} {\bibfnamefont {T.~J.~W.}\
  \bibnamefont {Wagner}}\ and\ \bibinfo {author} {\bibfnamefont
  {D.}~\bibnamefont {Vella}},\ }\Doi {10.1063/1.4724329} {\bibfield  {journal}
  {\bibinfo  {journal} {Applied Physics Letters},\ }\textbf {\bibinfo {volume}
  {100}},\ \bibinfo {pages} {233111} (\bibinfo {year} {2012})}\BibitemShut {NoStop}%
\bibitem [{\citenamefont {Cranford}(2013)}]{Cranford2013}%
  \BibitemOpen
  \bibfield  {author} {\bibinfo {author} {\bibfnamefont {S.~W.}\ \bibnamefont
  {Cranford}},\ }\Doi {10.1063/1.4788734} {\bibfield  {journal} {\bibinfo
  {journal} {Applied Physics Letters},\ }\textbf {\bibinfo {volume} {102}},\
  \bibinfo {pages} {031902} (\bibinfo {year} {2013})}\BibitemShut {NoStop}%
\bibitem [{\citenamefont {Malola}\ \emph {et~al.}(2008)\citenamefont {Malola},
  \citenamefont {H\"{a}kkinen},\ and\ \citenamefont
  {Koskinen}}]{malola_PRB_08}%
  \BibitemOpen
  \bibfield  {author} {\bibinfo {author} {\bibfnamefont {S.}~\bibnamefont
  {Malola}}, \bibinfo {author} {\bibfnamefont {H.}~\bibnamefont
  {H\"{a}kkinen}}, \ and\ \bibinfo {author} {\bibfnamefont {P.}~\bibnamefont
  {Koskinen}},\ }\href@noop {} {\bibfield  {journal} {\bibinfo  {journal}
  {Phys. Rev. B},\ }\textbf {\bibinfo {volume} {77}},\ \bibinfo {pages}
  {155412} (\bibinfo {year} {2008})}\BibitemShut {NoStop}%
\bibitem [{\citenamefont {Koskinen}(2012)}]{koskinen_PRB_12}%
  \BibitemOpen
  \bibfield  {author} {\bibinfo {author} {\bibfnamefont {P.}~\bibnamefont
  {Koskinen}},\ }\Doi {10.1103/PhysRevB.85.205429} {\bibfield  {journal}
  {\bibinfo  {journal} {Physical Review B},\ }\textbf {\bibinfo {volume}
  {85}},\ \bibinfo {pages} {205429} (\bibinfo {year} {2012})}\BibitemShut {NoStop}%
\bibitem [{\citenamefont {Zheng}\ \emph {et~al.}(2008)\citenamefont {Zheng},
  \citenamefont {Jiang}, \citenamefont {Liu}, \citenamefont {Weng},
  \citenamefont {Lu}, \citenamefont {Xue}, \citenamefont {Zhu}, \citenamefont
  {Jiang}, \citenamefont {Wang},\ and\ \citenamefont {Peng}}]{zheng_PRL_08}%
  \BibitemOpen
  \bibfield  {author} {\bibinfo {author} {\bibfnamefont {Q.}~\bibnamefont
  {Zheng}}, \bibinfo {author} {\bibfnamefont {B.}~\bibnamefont {Jiang}},
  \bibinfo {author} {\bibfnamefont {S.}~\bibnamefont {Liu}}, \bibinfo {author}
  {\bibfnamefont {Y.}~\bibnamefont {Weng}}, \bibinfo {author} {\bibfnamefont
  {L.}~\bibnamefont {Lu}}, \bibinfo {author} {\bibfnamefont {Q.}~\bibnamefont
  {Xue}}, \bibinfo {author} {\bibfnamefont {J.}~\bibnamefont {Zhu}}, \bibinfo
  {author} {\bibfnamefont {Q.}~\bibnamefont {Jiang}}, \bibinfo {author}
  {\bibfnamefont {S.}~\bibnamefont {Wang}}, \ and\ \bibinfo {author}
  {\bibfnamefont {L.}~\bibnamefont {Peng}},\ }\Doi
  {10.1103/PhysRevLett.100.067205} {\bibfield  {journal} {\bibinfo  {journal}
  {Physical Review Letters},\ }\textbf {\bibinfo {volume} {100}},\ \bibinfo
  {pages} {067205} (\bibinfo {year} {2008})}\BibitemShut {NoStop}%
\bibitem [{\citenamefont {Savini}\ \emph {et~al.}(2011)\citenamefont {Savini},
  \citenamefont {Dappe}, \citenamefont {\"{O}berg}, \citenamefont {Charlier},
  \citenamefont {Katsnelson},\ and\ \citenamefont
  {Fasolino}}]{savini_carbon_11}%
  \BibitemOpen
  \bibfield  {author} {\bibinfo {author} {\bibfnamefont {G.}~\bibnamefont
  {Savini}}, \bibinfo {author} {\bibfnamefont {Y.}~\bibnamefont {Dappe}},
  \bibinfo {author} {\bibfnamefont {S.}~\bibnamefont {\"{O}berg}}, \bibinfo
  {author} {\bibfnamefont {J.-C.}\ \bibnamefont {Charlier}}, \bibinfo {author}
  {\bibfnamefont {M.}~\bibnamefont {Katsnelson}}, \ and\ \bibinfo {author}
  {\bibfnamefont {A.}~\bibnamefont {Fasolino}},\ }\Doi
  {10.1016/j.carbon.2010.08.042} {\bibfield  {journal} {\bibinfo  {journal}
  {Carbon},\ }\textbf {\bibinfo {volume} {49}},\ \bibinfo {pages} {62}
  (\bibinfo {year} {2011})}\BibitemShut
  {NoStop}%
\bibitem [{foo()}]{footnote}%
  \BibitemOpen
  \href@noop {} {}\bibinfo {note} {With $9\sqrt{3\pi}\sqrt{\sqrt{7}-1}\approx
  35$}\BibitemShut {NoStop}%
\bibitem [{\citenamefont {{Das Sarma}}\ \emph {et~al.}(2011)\citenamefont {{Das
  Sarma}}, \citenamefont {Adam}, \citenamefont {Hwang},\ and\ \citenamefont
  {Rossi}}]{DasSarma2011}%
  \BibitemOpen
  \bibfield  {author} {\bibinfo {author} {\bibfnamefont {S.}~\bibnamefont {{Das
  Sarma}}}, \bibinfo {author} {\bibfnamefont {S.}~\bibnamefont {Adam}},
  \bibinfo {author} {\bibfnamefont {E.}~\bibnamefont {Hwang}}, \ and\ \bibinfo
  {author} {\bibfnamefont {E.}~\bibnamefont {Rossi}},\ }\Doi
  {10.1103/RevModPhys.83.407} {\bibfield  {journal} {\bibinfo  {journal}
  {Reviews of Modern Physics},\ }\textbf {\bibinfo {volume} {83}},\ \bibinfo
  {pages} {407} (\bibinfo {year} {2011})}\BibitemShut {NoStop}%
\bibitem [{\citenamefont {Wei}\ \emph {et~al.}(2012)\citenamefont {Wei},
  \citenamefont {Wu}, \citenamefont {Yin}, \citenamefont {Shi}, \citenamefont
  {Yang},\ and\ \citenamefont {Dresselhaus}}]{wei_nmat_12}%
  \BibitemOpen
  \bibfield  {author} {\bibinfo {author} {\bibfnamefont {Y.}~\bibnamefont
  {Wei}}, \bibinfo {author} {\bibfnamefont {J.}~\bibnamefont {Wu}}, \bibinfo
  {author} {\bibfnamefont {H.}~\bibnamefont {Yin}}, \bibinfo {author}
  {\bibfnamefont {X.}~\bibnamefont {Shi}}, \bibinfo {author} {\bibfnamefont
  {R.}~\bibnamefont {Yang}}, \ and\ \bibinfo {author} {\bibfnamefont
  {M.}~\bibnamefont {Dresselhaus}},\ }\Doi {10.1038/nmat3370} {\bibfield
  {journal} {\bibinfo  {journal} {Nature materials},\ }\textbf {\bibinfo
  {volume} {11}},\ \bibinfo {pages} {759} (\bibinfo {year} {2012})}\BibitemShut {NoStop}%
\bibitem [{\citenamefont {Zhu}\ \emph {et~al.}(2012)\citenamefont {Zhu},
  \citenamefont {Low}, \citenamefont {Perebeinos}, \citenamefont {Bol},
  \citenamefont {Zhu}, \citenamefont {Yan}, \citenamefont {Tersoff},\ and\
  \citenamefont {Avouris}}]{zhu_NL_12}%
  \BibitemOpen
  \bibfield  {author} {\bibinfo {author} {\bibfnamefont {W.}~\bibnamefont
  {Zhu}}, \bibinfo {author} {\bibfnamefont {T.}~\bibnamefont {Low}}, \bibinfo
  {author} {\bibfnamefont {V.}~\bibnamefont {Perebeinos}}, \bibinfo {author}
  {\bibfnamefont {A.~a.}\ \bibnamefont {Bol}}, \bibinfo {author} {\bibfnamefont
  {Y.}~\bibnamefont {Zhu}}, \bibinfo {author} {\bibfnamefont {H.}~\bibnamefont
  {Yan}}, \bibinfo {author} {\bibfnamefont {J.}~\bibnamefont {Tersoff}}, \ and\
  \bibinfo {author} {\bibfnamefont {P.}~\bibnamefont {Avouris}},\ }\Doi
  {10.1021/nl300563h} {\bibfield  {journal} {\bibinfo  {journal} {Nano
  letters},\ }\textbf {\bibinfo {volume} {12}},\ \bibinfo {pages} {3431}
  (\bibinfo {year} {2012})}\BibitemShut
  {NoStop}%
\bibitem [{\citenamefont {Duong}\ \emph {et~al.}(2012)\citenamefont {Duong},
  \citenamefont {Han}, \citenamefont {Lee}, \citenamefont {Gunes},
  \citenamefont {Kim}, \citenamefont {Kim}, \citenamefont {Kim}, \citenamefont
  {Ta}, \citenamefont {So}, \citenamefont {Yoon}, \citenamefont {Chae},
  \citenamefont {Jo}, \citenamefont {Park}, \citenamefont {Chae}, \citenamefont
  {Lim}, \citenamefont {Choi},\ and\ \citenamefont {Lee}}]{duong_nature_12}%
  \BibitemOpen
  \bibfield  {author} {\bibinfo {author} {\bibfnamefont {D.~L.}\ \bibnamefont
  {Duong}}, \bibinfo {author} {\bibfnamefont {G.~H.}\ \bibnamefont {Han}},
  \bibinfo {author} {\bibfnamefont {S.~M.}\ \bibnamefont {Lee}}, \bibinfo
  {author} {\bibfnamefont {F.}~\bibnamefont {Gunes}}, \bibinfo {author}
  {\bibfnamefont {E.~S.}\ \bibnamefont {Kim}}, \bibinfo {author} {\bibfnamefont
  {S.~T.}\ \bibnamefont {Kim}}, \bibinfo {author} {\bibfnamefont
  {H.}~\bibnamefont {Kim}}, \bibinfo {author} {\bibfnamefont {Q.~H.}\
  \bibnamefont {Ta}}, \bibinfo {author} {\bibfnamefont {K.~P.}\ \bibnamefont
  {So}}, \bibinfo {author} {\bibfnamefont {S.~J.}\ \bibnamefont {Yoon}},
  \bibinfo {author} {\bibfnamefont {S.~J.}\ \bibnamefont {Chae}}, \bibinfo
  {author} {\bibfnamefont {Y.~W.}\ \bibnamefont {Jo}}, \bibinfo {author}
  {\bibfnamefont {M.~H.}\ \bibnamefont {Park}}, \bibinfo {author}
  {\bibfnamefont {S.~H.}\ \bibnamefont {Chae}}, \bibinfo {author}
  {\bibfnamefont {S.~C.}\ \bibnamefont {Lim}}, \bibinfo {author} {\bibfnamefont
  {J.~Y.}\ \bibnamefont {Choi}}, \ and\ \bibinfo {author} {\bibfnamefont
  {Y.~H.}\ \bibnamefont {Lee}},\ }\Doi {10.1038/nature11562} {\bibfield
  {journal} {\bibinfo  {journal} {Nature},\ }\textbf {\bibinfo {volume}
  {490}},\ \bibinfo {pages} {235} (\bibinfo {year} {2012})}\BibitemShut {NoStop}%
\bibitem [{\citenamefont {Coleman}\ \emph {et~al.}(2011)\citenamefont
  {Coleman}, \citenamefont {Lotya}, \citenamefont {O'Neill}, \citenamefont
  {Bergin}, \citenamefont {King}, \citenamefont {Khan}, \citenamefont {Young},
  \citenamefont {Gaucher}, \citenamefont {De}, \citenamefont {Smith},
  \citenamefont {Shvets}, \citenamefont {Arora}, \citenamefont {Stanton},
  \citenamefont {Kim}, \citenamefont {Lee}, \citenamefont {Kim}, \citenamefont
  {Duesberg}, \citenamefont {Hallam}, \citenamefont {Boland}, \citenamefont
  {Wang}, \citenamefont {Donegan}, \citenamefont {Grunlan}, \citenamefont
  {Moriarty}, \citenamefont {Shmeliov}, \citenamefont {Nicholls}, \citenamefont
  {Perkins}, \citenamefont {Grieveson}, \citenamefont {Theuwissen},
  \citenamefont {McComb}, \citenamefont {Nellist},\ and\ \citenamefont
  {Nicolosi}}]{coleman_science_11}%
  \BibitemOpen
  \bibfield  {author} {\bibinfo {author} {\bibfnamefont {J.~N.}\ \bibnamefont
  {Coleman}}, \bibinfo {author} {\bibfnamefont {M.}~\bibnamefont {Lotya}},
  \bibinfo {author} {\bibfnamefont {A.}~\bibnamefont {O'Neill}}, \bibinfo
  {author} {\bibfnamefont {S.~D.}\ \bibnamefont {Bergin}}, \bibinfo {author}
  {\bibfnamefont {P.~J.}\ \bibnamefont {King}}, \bibinfo {author}
  {\bibfnamefont {U.}~\bibnamefont {Khan}}, \bibinfo {author} {\bibfnamefont
  {K.}~\bibnamefont {Young}}, \bibinfo {author} {\bibfnamefont
  {A.}~\bibnamefont {Gaucher}}, \bibinfo {author} {\bibfnamefont
  {S.}~\bibnamefont {De}}, \bibinfo {author} {\bibfnamefont {R.~J.}\
  \bibnamefont {Smith}}, \bibinfo {author} {\bibfnamefont {I.~V.}\ \bibnamefont
  {Shvets}}, \bibinfo {author} {\bibfnamefont {S.~K.}\ \bibnamefont {Arora}},
  \bibinfo {author} {\bibfnamefont {G.}~\bibnamefont {Stanton}}, \bibinfo
  {author} {\bibfnamefont {H.-Y.}\ \bibnamefont {Kim}}, \bibinfo {author}
  {\bibfnamefont {K.}~\bibnamefont {Lee}}, \bibinfo {author} {\bibfnamefont
  {G.~T.}\ \bibnamefont {Kim}}, \bibinfo {author} {\bibfnamefont {G.~S.}\
  \bibnamefont {Duesberg}}, \bibinfo {author} {\bibfnamefont {T.}~\bibnamefont
  {Hallam}}, \bibinfo {author} {\bibfnamefont {J.~J.}\ \bibnamefont {Boland}},
  \bibinfo {author} {\bibfnamefont {J.~J.}\ \bibnamefont {Wang}}, \bibinfo
  {author} {\bibfnamefont {J.~F.}\ \bibnamefont {Donegan}}, \bibinfo {author}
  {\bibfnamefont {J.~C.}\ \bibnamefont {Grunlan}}, \bibinfo {author}
  {\bibfnamefont {G.}~\bibnamefont {Moriarty}}, \bibinfo {author}
  {\bibfnamefont {A.}~\bibnamefont {Shmeliov}}, \bibinfo {author}
  {\bibfnamefont {R.~J.}\ \bibnamefont {Nicholls}}, \bibinfo {author}
  {\bibfnamefont {J.~M.}\ \bibnamefont {Perkins}}, \bibinfo {author}
  {\bibfnamefont {E.~M.}\ \bibnamefont {Grieveson}}, \bibinfo {author}
  {\bibfnamefont {K.}~\bibnamefont {Theuwissen}}, \bibinfo {author}
  {\bibfnamefont {D.~W.}\ \bibnamefont {McComb}}, \bibinfo {author}
  {\bibfnamefont {P.~D.}\ \bibnamefont {Nellist}}, \ and\ \bibinfo {author}
  {\bibfnamefont {V.}~\bibnamefont {Nicolosi}},\ }\Doi
  {10.1126/science.1194975} {\bibfield  {journal} {\bibinfo  {journal}
  {Science},\ }\textbf {\bibinfo {volume} {331}},\ \bibinfo {pages} {568}
  (\bibinfo {year} {2011})}\BibitemShut
  {NoStop}%
\bibitem [{\citenamefont {Plimpton}(1995)}]{plimpton_JCP_95}%
  \BibitemOpen
  \bibfield  {author} {\bibinfo {author} {\bibfnamefont {S.}~\bibnamefont
  {Plimpton}},\ }\href@noop {} {\bibfield  {journal} {\bibinfo  {journal} {J.
  Comp. Phys.},\ }\textbf {\bibinfo {volume} {117}},\ \bibinfo {pages} {1}
  (\bibinfo {year} {1995})}\BibitemShut {NoStop}%
\bibitem [{\citenamefont {Stuart}\ \emph {et~al.}(2000)\citenamefont {Stuart},
  \citenamefont {Tutein},\ and\ \citenamefont {Harrison}}]{stuart_JCP_00}%
  \BibitemOpen
  \bibfield  {author} {\bibinfo {author} {\bibfnamefont {S.~J.}\ \bibnamefont
  {Stuart}}, \bibinfo {author} {\bibfnamefont {A.~B.}\ \bibnamefont {Tutein}},
  \ and\ \bibinfo {author} {\bibfnamefont {J.~A.}\ \bibnamefont {Harrison}},\
  }\href@noop {} {\bibfield  {journal} {\bibinfo  {journal} {J. Chem. Phys.},\
  }\textbf {\bibinfo {volume} {112}},\ \bibinfo {pages} {6472} (\bibinfo {year}
  {2000})}\BibitemShut {NoStop}%
\bibitem [{\citenamefont {Dumitrica}\ and\ \citenamefont
  {James}(2007)}]{dumitrica_JMPS_07}%
  \BibitemOpen
  \bibfield  {author} {\bibinfo {author} {\bibfnamefont {T.}~\bibnamefont
  {Dumitrica}}\ and\ \bibinfo {author} {\bibfnamefont {R.~D.}\ \bibnamefont
  {James}},\ }\href@noop {} {\bibfield  {journal} {\bibinfo  {journal} {J.
  Mech. Phys. Solids},\ }\textbf {\bibinfo {volume} {55}},\ \bibinfo {pages}
  {2206} (\bibinfo {year} {2007})}\BibitemShut {NoStop}%
\bibitem [{\citenamefont {Koskinen}\ and\ \citenamefont
  {Kit}(2010)}]{koskinen_PRL_10}%
  \BibitemOpen
  \bibfield  {author} {\bibinfo {author} {\bibfnamefont {P.}~\bibnamefont
  {Koskinen}}\ and\ \bibinfo {author} {\bibfnamefont {O.~O.}\ \bibnamefont
  {Kit}},\ }\href@noop {} {\bibfield  {journal} {\bibinfo  {journal} {Phys.
  Rev. Lett.},\ }\textbf {\bibinfo {volume} {105}},\ \bibinfo {pages} {106401}
  (\bibinfo {year} {2010})}\BibitemShut {NoStop}%
\end{thebibliography}

%merlin.mbs 2010-03-15 4.21a (PWD, AO, DPC)
%Control: key (0)
%Control: author (72) initials jnrlst
%Control: editor formatted (1) identically to author
%Control: production of article title (-1) disabled
%Control: page (0) single
%Control: year (1) truncated
%Control: production of eprint (0) enabled
%

\end{document}